\documentclass[a4paper,11pt]{article}%
\newcommand{\AuthorA}{Paula~Sanz Leon}%
\newcommand{\AuthorB}{Ivo~Vanzetta}%
\newcommand{\AuthorC}{Guillaume S.~Masson} %
\newcommand{\AuthorD}{Laurent U.~Perrinet}%
\newcommand{\AddressA}{INCM, CNRS / Aix-Marseille University, France}
\newcommand{\AddressB}{Institut des Neurosciences de la Timone, CNRS / Aix-Marseille University, Marseille, France}%
\newcommand{\Website}{http://invibe.net/LaurentPerrinet}%
\newcommand{\EmailA}{Paula.Sanz@univ-amu.fr}%
\newcommand{\EmailB}{Ivo.Vanzetta@univ-amu.fr}%
\newcommand{\EmailC}{Guillaume.Masson@univ-amu.fr}%
\newcommand{\EmailD}{Laurent.Perrinet@univ-amu.fr}%
\newcommand{\Title}{%
Motion Clouds: %
Model-based stimulus synthesis of natural-like random textures for the study of motion perception}%
\newcommand{\Abstract}{%
Choosing an appropriate set of stimuli is essential in order to characterize the response of a sensory system to a particular functional dimension, such as the eye movement following the motion of a visual scene. %
Here, we describe a framework to generate random texture movies with controlled information content, i.e., Motion Clouds. These stimuli are defined using a generative model which is based on controlled experimental parametrization. %
We show that Motion Clouds correspond to dense mixing of localized moving gratings with random positions. Their global envelope is similar to natural-like stimulation with an approximate full-field translation corresponding to a retinal slip. %
We describe the construction of these stimuli mathematically and propose an open-source python-based implementation. %
Examples of the use of this framework are shown. We also propose extensions to other modalities such as color vision, touch and audition.%
}%
\newcommand{\Keywords}%
{Motion Clouds; Random Phase Textures; Low-level sensory systems; eye movements; optimal stimulation; natural scenes; motion detection; random textures; Python}
\newcommand{\Acknowledgments}{%
This work is supported by the European Union project Number FP7-269921, ``BrainScaleS'' and by the VISAFIX (ANR-10-Blan-1432)  grant from the Agence Nationale de la Recherche. SLP is supported by a doctoral fellowship from the Minist{\`e}re de la Recherche. We thank Anna Montagnini, Fr\'ed\'eric Chavane and Gabriel Peyr\'e for their comments on an earlier version of the manuscript. Supplementary material can be accessed on the corresponding author's website at \textit{\Website/MotionClouds}.}

\usepackage[utf8x]{inputenc}
\usepackage[english]{babel}
\usepackage[pdftex]{graphicx} 
\usepackage{graphics} 
\usepackage{amssymb,amsmath,amsfonts}
\usepackage[usenames,dvipsnames]{color}
\usepackage[pdftex,colorlinks=true,linkcolor=black]{hyperref}
\usepackage{color}
\usepackage{units}
\usepackage{authblk}
\usepackage[comma,sort&compress,square]{natbib}

\DeclareGraphicsExtensions{.jpg,.pdf,.png,.tiff}
\hypersetup{
 unicode=false, 
	pdftitle={\Title},%
	pdfauthor={\AuthorA < \EmailA > \AddressA - \AuthorB < \EmailB > - \AuthorC < \EmailC > - \AuthorD < \EmailD > },
	pdfkeywords={\Keywords},%
	pdfsubject={\Acknowledgments}%
 pdfcreator={LaTeX}, 	
 pdfproducer={LUP,PSL}, 			
 colorlinks=true, 
 linkcolor = Black, 	
 citecolor = Black, 	
 filecolor = Black, 			 
 urlcolor = Black 		 
}

\title{\Title}%
\author[1,2]{\AuthorA \thanks{\EmailA}}
\author[1,2]{\AuthorB \thanks{\EmailB}}
\author[1,2]{\AuthorC \thanks{\EmailC}}
\author[1,2]{\AuthorD \thanks{Corresponding Author, \EmailD}}
\affil[1]{\AddressA}
\affil[2]{\AddressB}
\date{}
\begin{document}
\maketitle
\begin{abstract}
\Abstract
\end{abstract}%
\section*{Keywords}
\Keywords
\section*{Acknowledgments}
\Acknowledgments
\section{Introduction }\label{intro}

One of the objectives of system neuroscience is to understand how sensory information is encoded and represented in the central nervous system, from single neurons to population of cells forming columns, maps and large-scale networks. Unveiling how sensory-driven behaviors such as perception or action are elaborated implies to decipher the role of each processing stage, from peripheral sensory organs up to associative sensory cortical areas. There is a long tradition of probing each of these levels using standardized stimuli of low dimension and simple statistics. They are based on a powerful, but stringent theoretical approach that considers the visual system as a spatiotemporal frequency analyzer~\citep{Graham79,Watson83}. Accordingly, visual neurons have long been tested with drifting gratings in order to characterize both their selectivities and some of non-linear properties of their receptive fields~\citep{DeValois1988Spatial}. A similar approach was applied at both mesoscopic and macroscopic scales to define functional properties of cortical maps (e.g.~\citep{Blasdel86,TsO90}) and areas (e.g.~\citep{Singh2000Spatiotemporal,Henriksson08}), respectively.  

A more recent trend has been to consider sensory pathways as complex dynamical systems. As such, these are able to process high dimensional sensory inputs with complex statistics such as encountered during natural life. As a consequence, the objective is to understand how the visual brain encodes and processes natural visual scenes~\citep{Dan96}. This has led to new theoretical approaches of neuronal information processing~\citep{Field99}, as well as to the search for new sets of stimuli for measuring neuronal responses to complex sensory inputs (see~\citep{Touryan01,Wu06}). Controversial opinions have been proposed on whether natural scenes and movies should be used straightforwardly for visual stimulation as in~\citep{Felsen2005Natural} or whether one should rather develop new sets of ``artificial" stimuli. Importantly, the latter approach has the advantage of being relatively easy to parametrize and to customize at different spatial and temporal scales~\citep{Rust05}. In brief, it has become a critical challenge to elaborate new visual stimuli that fulfill these two constraints: being both efficient and relevant to probe high-dimension dynamical systems on the one hand, and on the other, being easily tailored so that they can be used to conduct quantitative experiments at different scales, from single neuron to behavior.
 
Here, our aim is to provide such a set of stimuli cast into a well-defined mathematical framework. We decided to focus on motion detection, as a good illustration for the search for optimal high-dimension stimuli. Visual motion processing is critically involved in several essential aspects of low- and middle vision such as scene segmentation, feature integration and object recognition (see~\citep{Braddick1993Segmentation,Bradley08,Burr11} for reviews). It also provides essential aspect of visual information for motor systems such as speed and direction of moving objects, as well as about self-motion. Lastly, it is one of the few systems for which an integrated approach from single neuronal activity to complex behaviors can be achieved using nearly identical experimental conditions, in order to elucidate the neural bases of perceptual decisions~\citep{Newsome1997Deciding} and motor responses (see~\citep{Masson10} for a collection of examples). 

However, motion perception is highly dynamical and the classical toolbox of standard motion stimuli (such as dots, bars, gratings or plaids) is now largely outdated and insufficient to understand how the primate brain achieves visual motion processing with both high efficiency and short computing time. To be optimal, a new set of stimuli should be rooted in theoretical assumptions about how motion information is processed~\citep{Watson95}. A large bulk of experimental and theoretical evidences support the view that local motion information is extracted through a set of spatiotemporal frequency analyzers, whose outputs are then integrated to yield motion direction and amplitude~\citep{Adelson85,Simoncelli98}. However, we still lack a deep understanding of several linear (L) and nonlinear (NL) operations needed to extract the global motion from the local luminance changes (see~\citep{Derrington04} for a recent review). For instance, it remains unclear how MT neurons can encode speed and direction independently of the local spatiotemporal frequency or orientation content of the image (see~\citep{Bradley08} for a recent review). It is also hard to predict MT neurons responses to dense noise patterns or natural scenes from their spatiotemporal frequency selectivity as explored with low-dimension stimuli~\citep{Priebe06,Nishimoto11}. Lastly, neuronal responses to natural movies are more reliable and sparse than when driven by low dimensional stimuli such as drifting gratings~\citep{Vinje00}. %

To overcome thee limits, several recent studies have proposed that linearly combining several frequency channels can partly account for pattern direction and speed selectivity~\citep{Rust2005Spatiotemporal,Rust06,Nishimoto11}. Still, such multistage L-NL models~\citep{Heeger96,Simoncelli98} fail to account most of the response properties seen with natural scenes (see~\citep{Carandini05} for a review). One key issue is to understand how motion information gathered at different scales is normalized and weighted before integration as in the divisive normalization version of the L-NL model of motion detection. Natural-like stimuli are good probes to further explore the performance of these models~\citep{Schwartz01}. However, ``raw" natural scenes have the major drawback that information content is poorly controlled: Their dimensionality is extremely high and the inter-stimulus variability in the information content with respect to sensory parameters is large~\citep{Rust05}. Popular alternatives to natural scenes are dense and sparse noise. However, those are often irrelevant to the sensory system and most often fail to drive strong neuronal responses.  Here, we explore a new approach for the characterization of the first-order motion system. Our stimuli are  equivalent to a sub-class of random phase textures (RPTs)~\citep{Galerne10}, which are increasingly attracting interest in exploring neural mechanisms of texture perception (e.g.~\citep{Solomon10})

The paper is organized as follows. In the Method section, we first recall the main properties of  RPTs as originally defined in computer vision for texture analysis. Next, we define their dynamical version, called thereafter Motion Clouds (MCs), and provide their complete mathematical formulation. We briefly describe the architecture of our implementation, all technical details being available as supplementary material, including the source code. In the Results section, we illustrate the practical use of MCs for studying several long-lasting problems of visual motion processing such as 2D motion integration, motion segmentation and transparency. For each, we will compare the usefulness of Motion Clouds relative to existing low-dimension stimuli. Finally, we discuss how this approach can be generalized to different aspects of visual system identification. %

\begin{figure}
	\includegraphics[width=\textwidth]{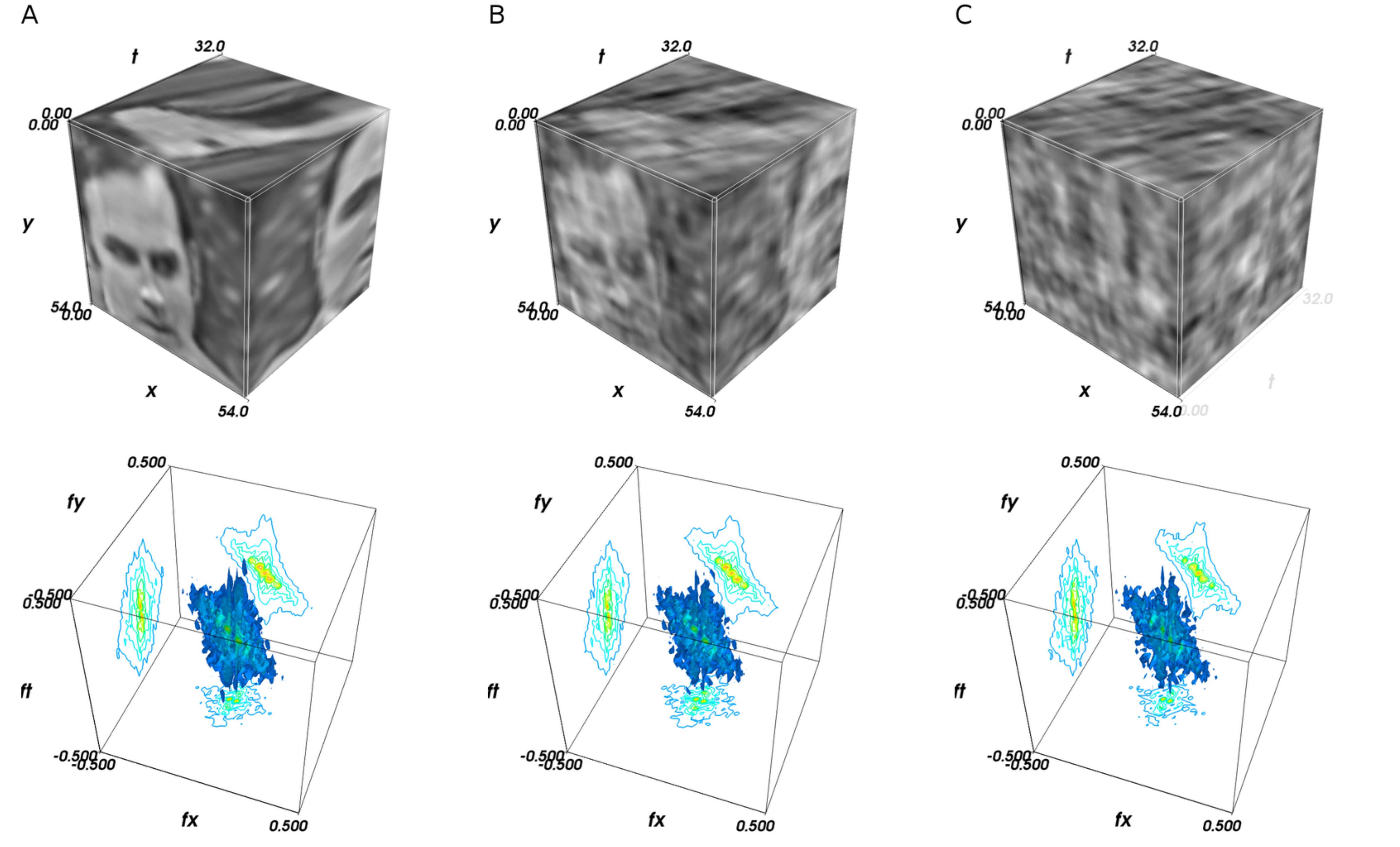}%
	\caption{(\textit{A}) Top: a natural movie with the main motion component consisting of a horizontal, rightward full-field translation. Such a movie would be produced by an eye movement with constant mean velocity to the left (negative $x$ direction), plus some residual, centered jitter noise in the motion-compensated natural scene. We represent the movie as a cube, whose $(x,y,t=0)$ face corresponds to the first frame, the $(x,y=0,t)$ face shows the rightward translation motion as diagonal stripes. As a result of the horizontal motion direction, the $(x=54,y,t)$ face is a reflected image of the $(x,y,t=0)$ face, contracted or dilated depending on the amplitude of motion. The bottom panel shows the corresponding Fourier energy spectrum, as well as its projections onto three orthogonal planes. For any given point in frequency space, the energy value with respect to the maximum is coded by 6 discrete color iso-surfaces (i.e.: 90\%, 75\%, 50\%, 25\%, 11\% and 6\% of peak. The amplitude of the Fourier energy spectrum has been normalized to 1 in all panels and the same conventions used here apply to all following figures. (\textit{B}) to (\textit{C}): The image is progressively morphed (A through B to C) into a Random Phase Texture by perturbing independently the phase of each Fourier component.  (\textit{upper row}): Form is gradually lost in this process, whereas (\textit{lower row}): most motion energy information is preserved, as it is concentrated around the same speed plane in all three cases (the spectral envelopes are nearly identical).}%
	 \label{fig:morphing}%
\end{figure}%

\section{Methods}\label{section:Methods}
\subsection{Random phase textures and natural retinal motion}\label{subsection:random_phase_textures}
First, Random Phase Textures (RPTs) are defined as generic random motion textures that are optimal for probing luminance-based visual processing. Most of the information present in a given dynamical image can be divided into its geometry (that is the outline of the objects it represents) and its distribution of luminance in space and time~\citep{Neri98,Perrone01,Perrone02,Jasinschi92}. In the spatiotemporal Fourier space this is well separated between the phase and the absolute amplitude spectra, respectively~\citep{Oppenheim81}. This can be easily seen by gradually perturbing the phase spectrum of a natural scene: while form is progressively lost, its global motion information remains essentially unchanged (see Figure~\ref{fig:morphing}). This invariance with respect to phase shuffling in the Fourier domain is generally considered to be characteristic of the first-order motion stage~\citep{Lu01,Derrington04}. 

\begin{figure}
	\includegraphics[width=\textwidth]{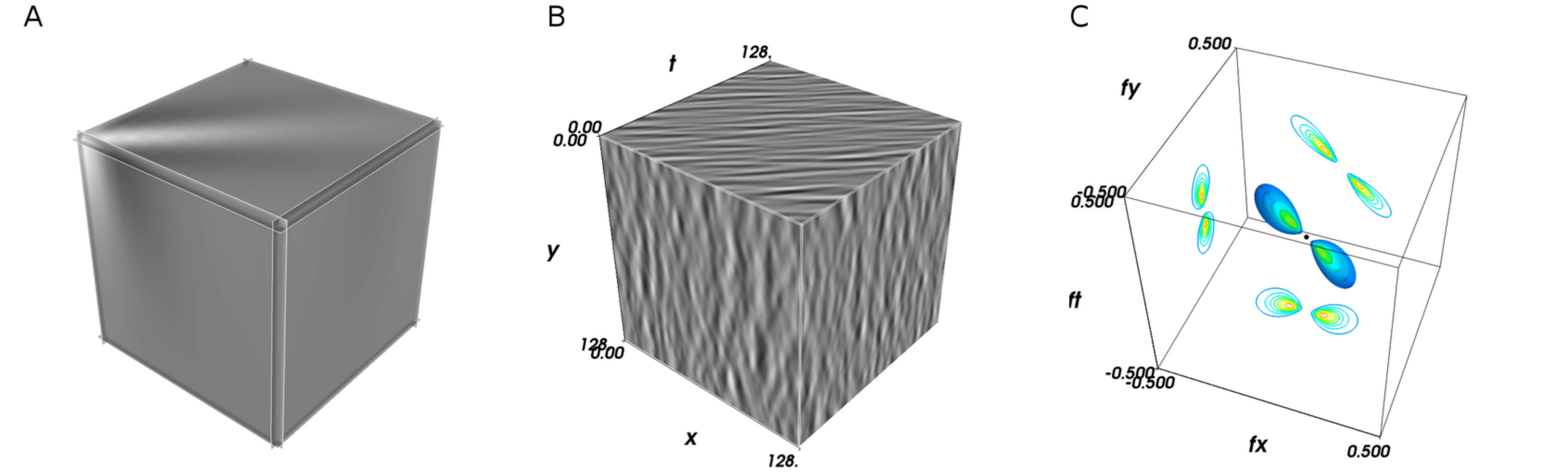}%
	 \caption{From an impulse to a Motion Cloud. (\textit{A}): The movie corresponding to a typical ``edge", i.e., a moving Gabor patch that corresponds to a localized grating. The Gabor patch being relatively small, for clarity, we zoomed 8 times into the non-zeros values of the image. (\textit{B}): By densely mixing multiple copies of the kernel shown in (A) at random positions, we obtain a Random Phase Texture (RPT), see Supplemental Movie 1. (\textit{C}):  We show here the envelope of the Fourier transform of kernel $K$: inversely, $K$ is the impulse response in image space of the filter defined by this envelope. Due to the linearity of the Fourier transform, apart from a multiplicative constant that vanishes by normalizing the energy of the  RPT to $1$, the spectral envelope of the RPT in (B) is the same as the one of the kernel K shown in (A): $\mathcal{E}_{\bar{\beta}}=\mathcal{F}(K)$. Note that, the spectral energy envelope  of a ``classical" grating would result in a pair of Dirac delta functions centered on the peak of the patches in (C) (the orange ``hot-spots"). Motion Clouds are defined as the subset of such RPTs whose main motion component is a full-field translations and thus characterized by spectral envelopes concentrated on a plane.}
	 \label{fig:impulse}
\end{figure}

We next formally define a linear generative model for the synthesis of such natural-like moving textures. Most generally, we can describe luminance at position $(x,y)$ and time $t$ as the scalar $I(x, y, t)$ that is the sum of the contribution of a set of basis functions:
\begin{equation}%
I(x, y, t) = \sum_k a_k \cdot G(x, y, t; \beta_k) \label{eq:gen1}%
\end{equation}%
The function $G$ defines the family of basis functions where each basis function is defined by parameters $\beta_k$. Scalars $a_k$ give the relative amplitude for each basis function and therefore will change for each individual image $I$, while the parameters $\beta_k$ are fixed for a set of stimuli. The advantage of this generative model is to separate the temporal scale of coding a specific moving stimulus (represented by the scalars $a_k$) from the temporal scale of a whole set of stimuli (as represented by the $\beta_k$). Efficient coding strategies use such generative models by optimizing scalars $a_k$ knowing a fixed set of basis functions $\beta_k$. 
Note that finding the optimal set $a_k$ knowing this linear generative model and the image $I$ is in general a non-linear problem (it is the \emph{coding} problem). When the set given by $\beta_k$ is large, this problem becomes difficult. %
In that context, divisive normalization gives a fair account for this problem using for its solution second-order correlations across basis functions~\citep{Schwartz01}. On a slower temporal scale, such model is used in neural modeling for studying the emergence of receptive fields by optimizing $\beta_k$, such as by using a Bayesian framework~\citep{Perrinet10shl}. %

In Fourier space, by linearity: 
\begin{equation}
\mathcal{F}(I)(f_x, f_y, f_t) = \sum_k a_k \cdot \mathcal{F}(G)(f_x, f_y, f_t; \beta_k) \label{eq:gen2}
\end{equation} 
where $\mathcal{F}$ is the Fourier transform. %
Here, we will use a fixed set of spatiotemporal Gabor kernels to implement localized, moving grating-like textures as they are known to efficiently code for natural images. For this particular set, parameters are defined as the peak's spatiotemporal position $\{x_k, y_k, t_k\}$, velocity (direction and speed), orientation and scale. We may write the translation of each component using the shift operator in the Fourier domain: 
\begin{equation}
\mathcal{F}(I)(f_x, f_y, f_t) = \sum_k a_k \cdot \mathcal{E}_{\bar{\beta_k}}(f_x, f_y, f_t) \cdot e^{-2i\pi (f_x \cdot x_k + f_y \cdot y_k + f_t \cdot t_k)} \label{eq:gen2bis}
\end{equation} 
where $\bar{\beta}_k$ denotes the parameters without positions $\{x_k, y_k, t_k\}$. In general, parameters $\bar{\beta}_k$ have some statistical regularities in Fourier space: For instance, velocity, orientation and scale parameters are correlated in space and time~\citep{Lewis84,Lagae09}. This defines an average spectral density envelope that we denote as $\mathcal{E}_{\bar{\beta}}$ and which is characteristic of the particular class of natural images that is coded~\citep{Torralba03}. %

We use this generative model to define RPTs and their Motion Clouds derivative that can be seen as first-order motion textures. If we shift randomly and independently the central position of edges (see Figure~\ref{fig:morphing}) and that this perturbation is stochastically independent from the distribution of the others parameters, one can describe the image by the following mean-field equation on its Fourier transform:
\begin{equation}%
\mathcal{F}(I)(f_x, f_y, f_t) = \mathcal{E}_{\bar{\beta}}(f_x, f_y, f_t) \cdot \sum_k a_k \cdot e^{-2i\pi (f_x \cdot x_k + f_y \cdot y_k + f_t \cdot t_k)} \label{eq:gen3}%
\end{equation}%
By consequence, the envelope is modulated by a stochastic spectrum that is defined at any point in Fourier space as the sum of random independent variables with the same distribution and variance. By virtue of the central limit theorem, we may define the set of stimuli $I$ as the random sequences generated by 1) an average envelope $\mathcal{E}_{\bar{\beta}}$, 2) a normally distributed, iid amplitude spectrum $A$, 3) a uniformly distributed phase spectrum $\Phi$ in $[0, 2\pi )$, that is to Random Phase Textures (RPT)~\citep{Galerne10} trivially extended to the spatiotemporal domain:
\begin{equation}%
\mathcal{F}(I) = \mathcal{E}_{\bar{\beta}} \cdot A \cdot e^{i\Phi} \label{eq:gen4}%
\end{equation}%

As noted by~\citep{Galerne10}, the main visual ingredients of RPTs are the envelope spectrum and the random phase spectrum, while $A$ has little perceptual effect. Indeed, removing the random amplitude spectrum, we still have a random fluctuation of the sign of each Fourier coefficient. From the central limit theorem, under the condition that the number of mixed components is large enough, the coefficient spectrum resulting from the mixing described by
\begin{equation}%
\mathcal{F}(I) = \mathcal{E}_{\bar{\beta}} \cdot e^{i\Phi} \label{eq:gen5}%
\end{equation}%
will still be given by a normal distribution and correspond to a correct generative model for RPTs as in Equation~\ref{eq:gen4}. Stimuli corresponding to such equations correspond to band-limited filtering of white-noise, that is, to a white noise (in space and time) linearly filtered by the kernel $K$, where $K=\mathcal{F}^{-1}(\mathcal{E}_{\bar{\beta}})$ corresponds to the average impulse response of the texture (see Figure~\ref{fig:impulse}). %

This class of random, textured, dynamical stimuli have several advantages over classical narrow bandwidth, low entropy stimuli, such as gratings or combinations of gratings. First, by varying the weight of each Fourier coefficient, we can vary its content and probe different types of motion integration models. Second, we can generate several different series of stimuli with different randomization seeds, while keeping all other parameters constant. Third, we can play with the bandwidth along each dimension to titrate the role of distributions of frequencies onto neuronal or behavioral responses. Fourth, we can reproduce the statistics of natural images by controlling the global envelope in Fourier space. Fifth, stochastic properties are generated only by varying the phase spectrum, without the need for adding noise component to the motion stimulus or controlling lifetime of individual features. Below we shall discuss several examples of the experimental usability of such stimuli. Stimuli similar to RPT have been already used. This was first formalized for the generation of natural-like static textures in computer vision~\citep{Lewis84} such as procedural or Perlin textures and is still largely used~\citep{Lagae09}. Mathematically, the resulting patterns are related to the morphogenesis studies pioneered by~\citep{Turing52}. Such static textures were used in psychophysics~\citep{Essock09}, in neurophysiology, for instance to study sensitivity of V1 neurons to dynamical expansion~\citep{Wang11} or nonlinear properties of non-classical receptive fields of primate MT neurons~\citep{Solomon10}. It is worth noting that a similar stimulus design was proposed for investigating another sensory system, i.e., audition~\citep{Rieke95,Klein00}. %

\subsection{Motion clouds as one particular type of random phase textures}
Defining optimal motion stimuli in order to probe the first-order, luminance-based motion system is a nontrivial problem. A straightforward approach is to generate static textures and to generate sequences as an exact, full-field translation of this static texture~\citep{Drewes08}. However, this approach is not generic enough. In particular, it lacks the possibility to vary the distribution of speeds being present in a given movie, a parameter that might be crucial to study precision and robustness of motion processing for perception or eye movements. Motion Clouds can be defined as the subset of RPTs that results from a generative model inspired by a rigid translation at central velocity of a large texture filling the whole visual field. This generative model will be specified by a central velocity $\vec{V}$ for the full-field translation, plus random independent perturbations of velocities around the central velocity, given by a bandwidth $B_V$. As a consequence, the spectral distribution of energy of such a sequence is centered on and squeezed onto a plane defined by the normal vector $\vec{V}$. The L-NL models of direction selectivity match the spatiotemporal properties of V1 or MT neuron receptive field with this plane. Phase information is concealed using the squared sum from the activity of receptive fields of odd and even phase~\citep{Adelson85}. By definition, Motion Clouds  using a similar envelope as given by the spatiotemporal filtering properties of V1 or MT neurons are thus equivalently defined as the set of stimuli that are optimally detected by these energy detectors~\citep{Nishimoto11}. Moreover, they are also optimal for motion coding in the information-theoretic sense, since they maximize entropy~\citep{Field94} compared to the presentation of a simple kernel $K$ as in~\citep{Watson95}. Similar random textures as Motion Clouds have been generated by displaying a rectangular grid of Gabor patches with random orientations and directions~\citep{Scarfe10}. However, such a regular grid introduces some geometrical information that may interfere with the processing of motion, as opposed to RPTs. %
Our Motion Clouds are more similar to the texture stimuli introduced by~\citep{Schrater00} or to the dynamical displays designed by~\citep{Tsuchiya07}. Below we propose one well-defined mathematical formalization for our Motion Clouds before presenting a solution for their implementation in psychophysical toolboxes. %

\section{Mathematical definition of Motion Clouds}\label{section:MathClouds}


We define Motion Clouds (MCs) as RPTs that are characterized by several key features. First, first-order motion information is  independent on changes in the phase of the Fourier coefficients of image sequences since it is contained in the amplitude of the spectral coefficients (Eq.~\ref{eq:gen4}).

\begin{equation}
\mathbf{I}(x, y, t) = \mathcal{F}^{-1}\left\{\mathcal{E}_{\bar{\beta}}(f_x,f_y,f_t) \cdot e^{i\phi(f_x,f_y,f_t)}\right\}
\label{eq:fourier_image}
\end{equation}
Second, in Fourier space, full-field, constant, translational motions correspond to an envelope $\mathcal{E}_{\bar{\beta}}$ whose distribution is concentrated on a speed  plane (a plane in Fourier space that contains the origin). Third, the distribution of the spectral envelope  $\mathcal{E}_{\bar{\beta}}$ is defined as a Gaussian. This is explained by the fact that Gabor filters have a Gaussian envelope and thus an optimal spread in Fourier space~\citep{Marcelja80}. As such they are well known models of simple cells in the primary visual cortex~\citep{Daugman80} that can describe the most salient properties of receptive fields and their tuning for spatial localization, orientation and spatial frequency selectivity (e.g.~\citep{Jones87}). Moreover,~\citet{Lee96}  derived the conditions under which a set of 2D Gabor wavelets are a suitable image basis for a complete representation of any image. This was further extended to the case of sequences with a known motion~\citep{Watson95} and therefore constitutes an accurate set for studying first-order motion. In summary, envelopes of MCs are essentially Gaussian distributions that are concentrated close to a speed plane (see Figure~\ref{fig:impulse}). Equivalently, these characteristics define MCs as {dense random mixing of spatiotemporal Gabor filters with similar speeds}. 

The implementation presented herein is based on a simplified parametrization of the envelope of the amplitude spectrum. Given that speed, radial frequency and orientation spread are independent, we can parametrize different types of MCs based on a factorization of each component. %
	
\begin{equation} %
\mathcal{E}_{\bar{\beta}}= \mathcal{V}_{(V_x, V_y, B_{V})} \times %
\mathcal{G}_{(f_0,B_{f})} \times \mathcal{O}_{(\theta,B_\theta)}  \times %
\mathcal{C}_{(\alpha)} %
\label{eq:synthesis} %
\end{equation} %
where all envelope parameters are given in their sub-label and envelopes correspond respectively to the speed plane, the frequency and the orientation tuning along this plane: %
\begin{enumerate}
\item For the speed envelope $\mathcal{V}$, two parameters define motion $\vec{V}= (V_x, V_y)$ (and thus the speed plane) while one parameter defines the bandwidth $B_{V}$ of this plane as we jitter the mean motion $\vec{V}$. Varying these parameters allows to study the response of motion detectors to different speeds and amounts of velocity noise. %
\item Projected onto the speed plane, we can define (i) the radial frequency envelope $\mathcal{G}$ with two parameters that set its mean value $f_0$ and bandwidth $B_{f}$.  Also (ii) the orientation envelope $\mathcal{O}$, is defined by two parameters: mean orientation $\theta$ and bandwidth $B_\theta$. In both cases, the two parameters can be thought as defining the nominal value and the uncertainty of each respective component of motion information. 
\item An additional envelope $ \mathcal{C}$ is parametrized by $\alpha$. It tunes the overall shape of the envelope similarly to what is observed in natural images.
\end{enumerate}

Note that one can modify the parameters of each envelope  independently and moreover, by the commutativity of the product operation, the order of the envelopes is arbitrary. It shall further be noticed that the actual values of each of these parameters can be set based on known properties of the biological system to be investigated, for each level of observation. We will now detail each of them. %

\subsection{Speed Envelope} %
The first axis of a Motion  Cloud is its speed component. Let us first recall that the Fourier transform of a static image with a global translation motion is the Fourier transform of the static image (that lies in the $f_t=0$ plane) tilted on a plane perpendicular to $\vec{V}=(V_x, V_y)$ defined as: %
\begin{equation} %
V_x\,f_x+V_y\,f_y + f_t = 0 %
\label{eq:speed_plane} %
\end{equation} %

The orientation and tilt of the plane are determined by the direction and speed of motion, respectively. For larger $V=\| \vec{V} \|=\sqrt{V_x^2+V_y^2}$, the tilt  becomes greater. To model speed variability (jitters) within a motion cloud, we shall assume that motion  varies slightly in both speed axes (i.e., direction and amplitude). Such envelope is given for instance by: %
\begin{equation} %
\mathcal{V}_{(V_x, V_y, B_{V})}(f_{x}, f_{y}, f_{t})=\exp\left(-\dfrac{1}{2}\left( \dfrac{ f_x\cdot V_x + f_y\cdot V_y + f_t }{B_{V} \cdot f_r} \right)^{2} \right)
\label{eq:envelope_speed_plane} %
\end{equation} %
where  $ f_r = \sqrt{f_x^2 + f_y^2 + f_t^2} $ is the radial frequency.  %

\subsection{Radial frequency envelope}
The second characteristic of a Motion  Cloud is its radial frequency envelope. This is defined as the one-dimensional distribution of radial frequency using spherical coordinates in the Fourier domain. Indeed, by spherical symmetry, this radial frequency envelope is then independent to motion and orientation tuning. An intuitive description of this envelope is a Gaussian distribution along this radial dimension, as it is often encountered to describe the frequency component of Gabor filters. An inconvenient of Gabor functions is the fact that their sum is not perfectly null. This shows up in Fourier space as a non-zero value at the origin. To overcome this issue we use the log-Gabor filters~\citep{Fischer07a}. A second advantage of using log-Gabor filters is that they better encode natural images~\citep{Field87}. We thus  build a spatial frequency band Gaussian filter that depends on the logarithm of the spatial radial frequency. We define $f_0$ as the mean radial frequency and $B_{f}$ as the filter's bandwidth. %

\begin{equation}
 \mathcal{G}_{(f_0,B_{f})} (f_{x}, f_{y}, f_{t})=\dfrac{1}{f_r}\,\exp\left(-\dfrac{1}{2}\left(\dfrac{\ln\left(\dfrac{f_r}{f_0}\right)}{\ln\left(\dfrac{f_0+B_{f}}{f_0}\right)}\right)^{2}\right)
\label{eq:envelope_sphere}
\end{equation}

\subsection{Orientation Envelope}
The third property of a Motion Cloud is its orientation envelope. Oriented structures in space-time yield oriented structures in the Fourier domain. Thus, the orientation component of the spectrum is given by the function $\mathcal{O} (f_x, f_y, f_t)$. It  is defined by a density function located at a mean orientation $\theta$ and whose spread is modeled using a Von-Mises distribution with parameter $B_{\theta}$ that represents its bandwidth centered on the symmetric with respect to the origin:
\begin{equation}
\mathcal{G} (f_x, f_y)= \exp \left( \dfrac{\cos(\theta_f-\theta)}{B_{\theta}}\right) + \exp \left( \dfrac{\cos(\theta_f-\theta-\pi)}{B_{\theta}}\right)
\label{eq:envelope_direction}
\end{equation}
where $\theta_f=\arctan (f_x, f_y)$ is the angle in the Fourier domain. Note again that this envelope is independent upon both speed and radial tuning. We define its bandwidth using the standard deviation $B_{\theta}$. If $B_{\theta}$ has a small value, a highly coherent orientation pattern is generated. %

\subsection{Spectral color} 
An important property of a Motion Clouds is their global statistics. Therefore, the average shape of its power spectrum must be controlled. It has been shown that the average power spectrum of natural scene follows a power law~\citep{Field87}:
\begin{equation} %
\mathcal{C}_{\alpha}(f_{x}, f_{y}, f_{t}) = \dfrac{1}{f_{\mathit{R}}^\alpha}
\label{eq:powerlaw} %
\end{equation} %
where $\alpha$  is usually set within the range  $0 < \alpha < 2$. By analogy with the color terminology used to characterize noise patterns, we call this function \textit{color envelope}. The simplest stimulus that can be built with our model is filtered spatial noise as a function of the power (exponent) factor $\alpha$.
In our model, we assume that the spectrum shape is independent of orientation and varies solely as a function of a radial frequency ($f_{\mathit{R}}$), defined as in~\citep{Schrater00} by:  %
\begin{equation} %
  f_{\mathit{R}} = \sqrt{f_{x}^{2} + f_{y}^{2} + \dfrac{f_{t}^{2}}{f_{t_0}^{2}}}
  \label{eq:freq_radius} %
\end{equation} %
The factor $f_{t_0}$ is a normalization factor and is associated to a normalized stimulus velocity. %

The color envelope weights the different frequency channels according to the statistics of natural images and therefore is optimal regarding the sensitivity of the primate visual system to the different spatiotemporal frequencies~\citep{Atick92}. In the examples given below,  we choose $\alpha = 1$, corresponding to a  a pink noise distribution. Note that this particular value allows for the marginal distribution integrated over all orientations to coincide with the speed and frequency envelope. Qualitatively, this global envelope does not change neither motion nor texture appearance of a Motion Cloud since it has no preferred speed, frequency or orientation. This is true in particular for Motion Clouds with a relatively narrow envelope in Fourier space. When using larger bandwidth values of the radial frequency distribution $B_{f}$, the shape of the global envelope becomes more important. %

\subsection{Implementation} \label{subsection:implementation} %
Since our objective is to provide a new set of stimuli for conducting neurophysiological or psychophysical experiments, we must propose a framework for generating and displaying Motion Clouds under well controlled parameter settings. Using standard computer libraries,  the theoretical framework described above can be implemented while taking into account technical constraints such as discretization and videographic displays. In the supplementary material, we provide with the source code used to generate our calibrated motion clouds using Python libraries. %

\section{Results}\label{section:Results} %

In order to illustrate how Motion Clouds can be used to investigate different aspects of motion processing, we now describe some of their applications. We emphasize how classical stimuli, such as gratings or plaid patterns, can be conveniently represented as Motion Clouds. This last aspect is important: Motion Clouds can be seen as a single class of motion stimuli encompassing both low-dimension and complex dynamical stimuli. It becomes then possible to parametrically investigate the effects of spatiotemporal frequency content upon different stages of motion processing. It shall be noticed that all the following examples are chosen such as to fit the characteristics of visual motion systems; yet, the same logic applies to other aspects of visual processing, such as texture or shape perception. %

\subsection{Motion Clouds equivalents of classical stimuli}
Sinusoidal luminance gratings are defined by a small set of parameters (orientation, direction, frequency). This translates naturally into a set of Motion Clouds with the parameters that we defined: speed, orientation, frequency. In addition, we now have the choice of 3 extra parameters $B_{V}$, $B_\theta$, $B_{f}$ that tune the bandwidth along each of these components, respectively (see Figure~\ref{fig:examples}-left). It thus becomes possible to investigate spatial or temporal frequency, orientation or direction selectivity, as well as the role of their respective tuning bandwidths. %

With drifting gratings, perceived motion direction is necessarily defined perpendicular to its orientation. This is related to the aperture problem: translation of an 1D elongated edge is ambiguous and its visual motion is compatible with an infinite number of velocity vectors~\citep{Movshon85}. A novel formulation of this problem can be designed by creating a Motion Cloud whose direction is not perpendicular to the main orientation and whose orientation bandwidth is very narrow. Indeed, a classical motion detector would then be incapable of determining non-ambiguously the speed plane that corresponds to such an envelope (see Figure~\ref{fig:examples}-middle). %

Motion Clouds also encompass textures similar to Random-Dot Kinetograms (RDKs). Usually, RDKs consist in a set of small dots drifting in a given direction and speed, each dot having a limited life time. This is similar to our original definition of Motion Clouds. Such pattern is defined in Equation~\ref{eq:gen1} with a kernel that would correspond to a Dirac delta function in space, a square ON and OFF function in time and a sparse set of coefficients $a_i$. Note that this kernel would correspond to a flat envelope on the speed plane with a bandwidth proportional to the inverse of the life-time of dots. This is therefore controlled in Motion Clouds by the parameter $B_{V}$ and indeed, we observe that shorter values induced ``features'' which last longer. We stress, however, that Motion Clouds are necessarily equivalent to dense, not sparse, noise patterns. %

Moreover, each Motion Cloud is generated by a fully known, computer-generated noise. It is therefore possible to regenerate exactly the same stimulus by using the same seed in the random number generator. This property allows to investigate inter-trial variability and thus the relative importance, for the system at hand, of external noise (measurement noise) and internal noise (uncertainty due to ambiguities and mixtures in the signal representation). This approach corresponds to the use of frozen noise stimuli~\citep{Mainen95}, that is, with a set of inputs for the visual system that are randomly generated but can be presented many times in a strictly identical manner. %
\subsection{Comparing broad band and narrow band motion stimuli}

\begin{figure}
\includegraphics[width=\textwidth]{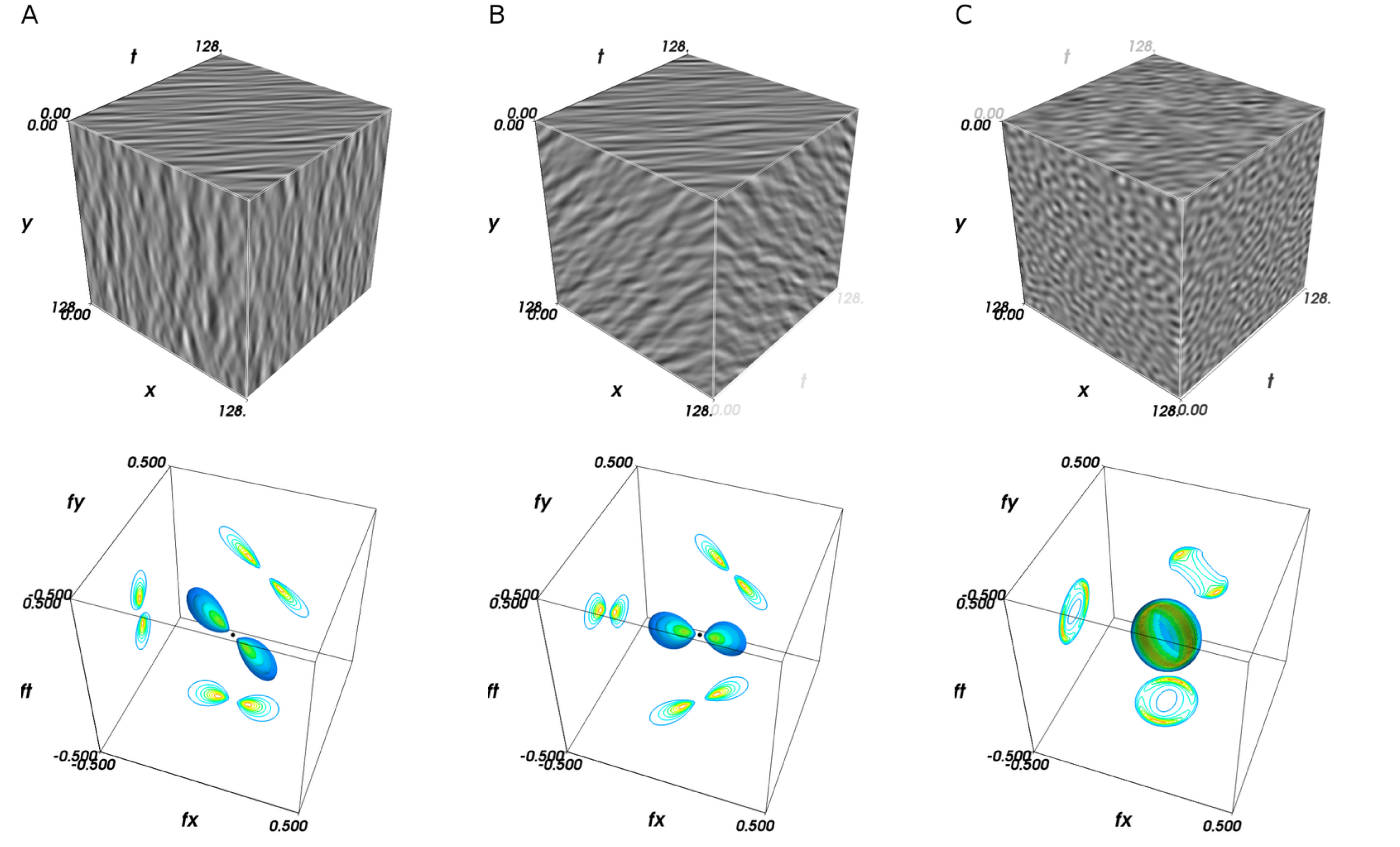}%
	 \caption{Equivalent MC representations of some classical stimuli. (\textit{A}, \textit{top}): a narrow-orientation-bandwidth Motion Cloud produced only with vertically oriented kernels and a horizontal mean motion to the right (Supplemental Movie 1). (\textit{Bottom}): The spectral envelopes concentrated on a pair of patches centered on a constant speed surface. Note that this ``speed plane" is thin (as seen by the projection onto the ($f_x$,$f_t$) face), yet it has a finite thickness, resulting in small, local, jittering motion components. ({\textit{B}}) a Motion Cloud illustrating the aperture problem. (\textit{Top}): The stimulus, having oblique preferred orientation ($\theta=\frac{\pi}{4}$ and narrow bandwidth $B_{\theta}=\pi/36$) is moving horizontally and rightwards. However, the perceived speed direction in such a case is biased towards the oblique downwards, i.e., orthogonal to the orientation, consistently with the fact that the best speed plane is ambiguous to detect (Supplemental Movie 2). (\textit{C}): a low-coherence random-dot kinematogram-like Motion Cloud: its orientation and speed bandwidths, $B_{\theta}$ and $B_{V}$ respectively, are large, yielding a low-coherence stimulus in which no edges can be identified (Supplemental Movie 3).} %
	 \label{fig:examples} %
\end{figure}

Varying the spatiotemporal frequency distribution from a grating-like stimulus to complex random phase textures should be a powerful method for investigating neuronal selectivity and cortical maps of extra-striate areas. Motion Clouds (and other types of RPT patterns) shall be able to drive cortical neurons known for receiving converging inputs from several spatiotemporal frequency channels~\citep{Rust05}. We have considered the idea of generating stimuli to explore the effects of varying a single bandwidth parameter: $B_{f}$, while setting $B_{V}$ and $B_{\theta}$ to some fixed values (with relatively low values to get some precision along these components). We use the name BroadBand stimuli (BB) for the MCs with a large $B_{f}$, whereas NarrowBand stimuli (NB) are MCs characterized by small $B_{f}$ values. As illustrated in Figure~\ref{fig:examples}-middle, BB and NB clouds are both symmetric, airfoil-shaped volumes. However, the broadband envelope contains more frequency information than the narrow band one. Therefore it is thought to better represent natural images. %

\begin{figure}
\includegraphics[width=\textwidth]{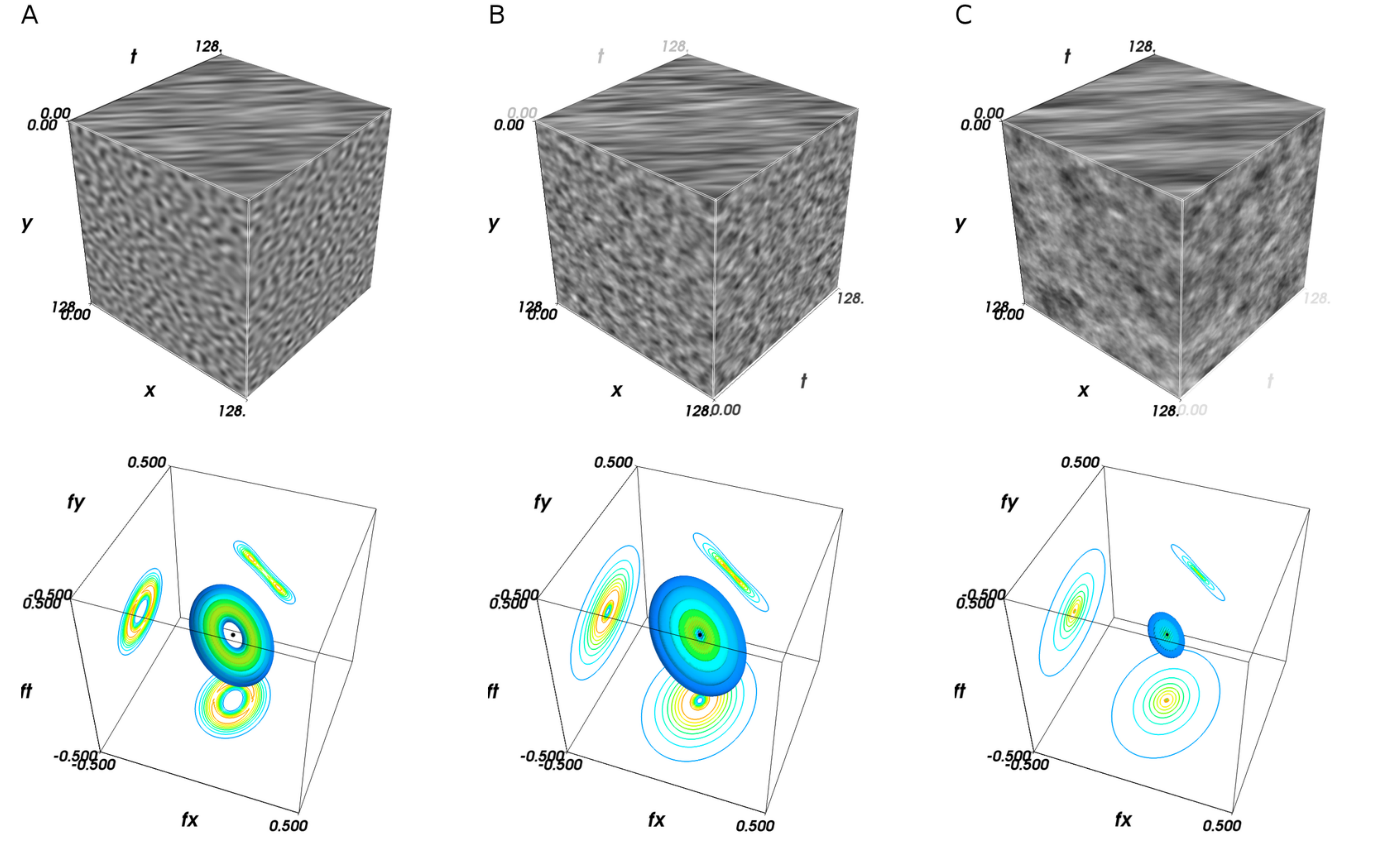}%
	 \caption{Broadband vs. narrowband stimuli. From (\textit{A}) through (\textit{B}) to (\textit{C}) the frequency bandwidth $B_{f}$ increases, while all other parameters (such as $f_{0}$) are kept constant. The Motion Cloud with the broadest bandwidth is thought to best represent natural stimuli, since, as those, it contains many frequency components. (\textit{A}) $B_{f}=0.05$ (Supplemental Movie 4), (\textit{B}) $B_{f}=0.15$ (Supplemental Movie 5) and (\textit{C}) $B_{f}=0.4$ (Supplemental Movie 6).} 
	 \label{fig:broadbandstimuli}
\end{figure}

Recently, we have used such Motion Cloud stimuli to investigate how the visual system integrates different spatial frequency information levels, by varying $B_{f}$ across a large range of spatial frequencies~\citep{Simoncini11}. The stimuli were displayed using Psychotoolbox v3~\citep{Pelli97,Brainard97} for MATLAB (http://psychtoolbox.org) on a CRT monitor ($1280\times1024$@\unit[100]{Hz}). They covered \unit[47]{degrees} of visual angle at a viewing distance of \unit[57]{cm}. We have used these stimuli to understand how two different visual behaviors, perceptual speed discrimination and reflexive tracking, would take advantage of presenting a single speed at different spatial scales. We found that the visual system pools motion information adaptively, as a function of constraints raised by different tasks. Motion Clouds were found to be useful to resolve problems associated with the integration of multiple spatial frequencies as they allow a precise control all variables related to speed and frequency content. In particular, previous studies have failed to understand how speed information is reconstructed across different spatial frequencies because the mixing of two, or more, gratings poses several perceptual problems. For instance, depending on the phase relationship between spatial frequency components, different interference patterns would appear, generating second-order motion in same or opposite motion direction~\citep{smith90}. Second, mixing sparse RDKs moving at the same speed but band-pass filtered at different spatial frequency results in complex patterns that can be perceived as being either coherent or transparent~\citep{watson94}. The same difficulties have been encountered by neurophysiological studies trying to understand the origin of speed selectivity in V1 complex cells~\citep{Priebe06} or MT neurons~\citep{Priebe03}. %

\subsection{Clouds with competing motions}%
Low and mid-level visual integration and segmentation mechanisms have been extensively investigated with either combinations of gratings (i.e. plaid patterns) or random dot patterns with different directions, speed and/or spatiotemporal components. Such plaid stimuli have been extensively studied and constitute an important pillar in motion detection theories, such as the separation between component and pattern cells in area MT (see~\citep{Movshon85,Burr11,Bradley05} for reviews). However, there have been a long standing controversy about which information can be used in plaid patterns, such as component gratings, their product or 2D features called blobs that are generated at the intersection between component gratings~\citep{Derrington04}. It is also unclear how different direction and spatial frequency channels are mixed to create pattern direction selectivity~\citep{Rust06}. As explained above, Motion Clouds stimuli are by definition less susceptible to create interference patterns (or Moir{\'e} patterns) when mixed together. This is a striking difference with respect to classical low-entropy stimuli, such as gratings. Being able to mix together two textures with the same motion but different characteristic spatial frequencies is also critical to further study motion integration (e.g. single neuron selectivity:~\citep{Rust05}; ocular tracking behavior:~\citep{Masson02}; motion perception~\citep{smith90}). By contrast, it must be also possible to mix two textures with different motions to study the competition between integration and segmentation, leading to different percepts such as coherent or transparent motions. %

\begin{figure}
\includegraphics[width=\textwidth]{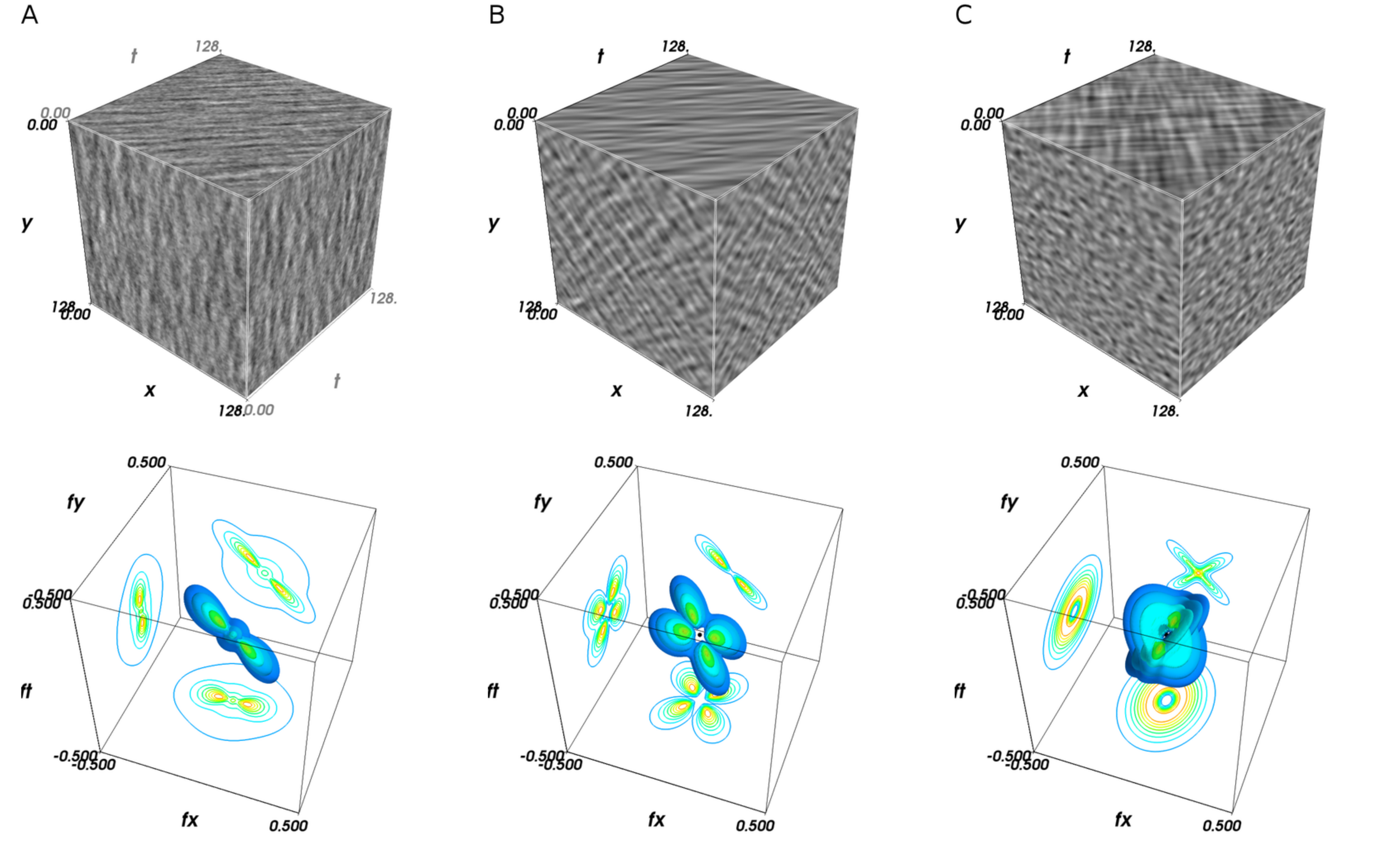}%
	 \caption{Competing Motion Clouds. (\textit{A}): A narrow-orientation-bandwidth Motion Cloud with explicit noise. A red noise envelope was added to the global envelop of a Motion Cloud with a bandwidth in the orientation domain (Supplemental Movie  7). (\textit{B}): Two Motion Clouds with same motion but different preferred orientation were added together, yielding a plaid-like Motion Cloud texture (Supplemental Movie 8). (\textit{C}): Two Motion Clouds with opposite velocity directions were added, yielding a texture similar to a ``counter-phase'' grating (Supplemental Movie 9). Note that the crossed shape in the $f_x-f_t$ plane is a signature of the opposite velocity directions, while two gratings with the same spatial frequency and in opposite directions would generate a flickering stimulus with energy concentrated on the $f_t$ plane.} %
	 \label{fig:counterphase} %
\end{figure} %
Using Motion Clouds, there is a further number of combinations that may be of interest for studying motion detection. We illustrate several possibilities in Figure~\ref{fig:counterphase}. The left panel shows a standard Motion Cloud with added explicit noise, corresponding to an envelope broadly centered around $V=0$. The middle panel illustrates the plaid-equivalent Motion Cloud obtained by adding two Motion Clouds of same velocity but different orientations, similarly to a plaid stimulus. In the right panel, the two components have different velocities (here opposite ones) while all other parameters are identical. With standard gratings, such two gratings would interfere and create a {counter-phase}, flickering stimulus. With Motion Clouds, there is no such interference and the resulting stimulus has all desired energy distributed on both speed planes. By varying the relative direction of two, or more, components, it becomes possible to produce several transparent patterns and therefore to overcome a limit of classical motion stimuli such as gratings. %

\section{Discussion}\label{section:future}

In this article we described the mathematical framework and provide the computer implementation of a set of complex stimuli that we call Motion Clouds. Those  are an instantiation of a more generic class of stimuli called Random Phase Textures.  These stimuli, presented herein in the context of visual motion perception, represent an attempt to fill the gap between simple stimuli (such as spots of light or sinusoidal gratings), stimulus ensembles consisting of simple stimuli (for instance, white noise patterns) and natural stimuli~\citep{Felsen2005Natural,Rust05}. Similar approaches have been  used before in the case of motion detection~\citep{Schrater00} but stimuli have been described in a somewhat incomplete and non accessible way. Here, our goal was to provide a complete and rigorous  mathematical description of those stimuli, as well as tools for generating them. We have also given a few examples of different subset of Motion Clouds that could be used for probing detection, integration and segmentation stages at both psychophysical and neurophysiological levels. To conclude, we indicate a few future extensions and possible uses. %
\subsection{Embedding spectral properties of natural images}
Both sensory and motor functions are natural tasks and therefore it is essential to understand how they deal with natural stimuli. Following the core principles of Natural Systems Analysis~\citep{Geisler08}, we think it is possible to extend our model of natural stimulation to carry out different and more complex experiments in the visual system. Seminal work from~\citep{Zeki83} showed that there exists a selectivity for color in higher visual areas such as V4 of the macaque monkey. Moreover,  the work by~\citep{Conway07}  shows that in the extra-striate cortex (V3, V4, Inferior Temporal Cortex) color is processed in terms of the full range of hues found in color space. The spatial structure is represented by  'globs' which are clustered by color preference, and organized as color columns. It is therefore important to develop an extension of MCs to be able to probe color vision. A first approach would consist in  creating a simple colored MC using an RGB scheme~\citep{Galerne10}. In this case we should add the same uniform random phase to each color channel. More realistically, a short medium long-cone (SML) scheme will have to be used, taking into account the cone fundamentals. Such color texture stimuli would permit to create a wide variety of new psychophysics experiments related to color perception. %
\subsection{Exploiting phase parameters: towards a systematic exploration of the role of geometry }
The amplitude spectra of natural images are characterized by their \unitfrac{1}{f} shape; in consequence, the global power spectrum cannot provide much information on any natural image that can be used, for instance, for fine pattern recognition or classification  (e.g.~\citep{VictorComte1996}, see also~\citep{Oliva01,Torralba03}). Information contained within the phase spectrum is therefore the key to {identifying} the contents in the images, i.e, how shape is coded in natural images. This implies that the visual system must be sensitive to the phase structure of artificial stimuli or natural images, at least at some spatial scales~\citep{PhillipsTodd2010,HansenHess2006}. This could be related to the rich representation of phase provided by the receptive field structure of visual neurons, from primary visual cortex up to extra-striate areas and further. We believe that Motion Clouds --and to a larger extent Random Phase Textures--- are powerful tools to probe the properties of phase-sensitive mechanisms in neural populations. In the cases presented above, patterns have parametrized random phases: phase values are drawn from a uniform probability distribution. However, one can evidently draw these phases according to some structure known \textit{a priori}, for instance, by correlating the phase of edges with similar orientations. This would progressively introduce collinearities in the set of stimuli, as needed to trigger short-range properties within the so-called association fields~\citep{hessreview03}. By manipulating these parameters, we shall be able to control for the detailed information content in the different axes of the corresponding associative field, for instance the role of collinearity versus cocircularity~\citep{Perrinet11sfn}. %
\subsection{Extension to increased complexity}
Random textured dynamical stimuli are generated as instances of few random variables defined by a generative model of synthesis. As a consequence, one may control the structural complexity of these synthetic textures by tuning the structure of the generative equations. In fact, the geometry of the visual world can be handled by using models to deal directly with the statistics of concurrent parameters, for instance edges or textures. For  example, within each texture and/or edge class, low-dimensional models control the complexity of the stimuli using few meaningful inputs (regularity of edges, number of crossings, curvature of the texture flow, etc). This complexity parametrization gives access to both the local geometry of the image (for instance its local orientation, frequency, scale, granularity) and to more global integration properties (good continuation of edges, approximate periodicity). %

These models can be assembled, thus leading to a rich content that mixes edge and texture patterns. It is believed that this hierarchical structure of generative models maps on a one-to-one basis with the structure of the visual system, from the detection of moving contrasts in the retina through edges in the primary visual cortex up to higher order attributes like motion and shape.  By designing such models with increasing scales of complexity, it shall therefore be possible to specifically target structures in the low-level visual system, such as respectively V1, V2, V4 and MT. The generative framework underlying Motion Clouds can make an important contribution to this long-term goal. %
\subsection{Exporting Motion Clouds to other sensory modalities}
Strong parallels have been drawn between visual and haptic processing for low level encoding of motion information, for instance~\citep{Pei11}. Simple stimuli like drifting relief gratings, dynamic noise patterns or single elements such as lines and spots have already been used to investigate the properties of the somatosensory system. There is also a strong need to develop more sophisticated stimuli that can reproduce, in a controllable way, the statistics of natural somesthetic inputs.  The theoretical framework described in the present article may also be used to design somesthetic inputs using mechanical actuators to excite the vibrissal array of rats' whiskers~\citep{Jacob08}. This is another potential application of a set of stimuli bridging the gap between artificial and natural sensory input. %
%

\end{document}